\documentclass[aps,pra,floatfix,superscriptaddress,10pt]{revtex4-2}

\usepackage[utf8]{inputenc}
\usepackage[english]{babel}
\usepackage{graphicx}
\usepackage{amsmath,times}
\usepackage{amsfonts}
\usepackage{mathrsfs}
\usepackage{physics}
\usepackage[colorlinks=true,hyperfootnotes=true,breaklinks=true,citecolor=red,urlcolor=blue,linkcolor=blue]{hyperref}
\usepackage{mathbbol} 
\usepackage{ragged2e}
\usepackage{soul}

\usepackage{tikz}
\usetikzlibrary{shapes.geometric}

\usepackage{nicefrac}

\begin{document}

\title{Quantum Sensing of Defects}

\author{Simone Cavazzoni}
\email{simone.cavazzoni@kaist.ac.kr}
\affiliation{Department of Physics, Korea Advanced Institute of Science and Technology, Daejeon 34141, Korea}%
\author{Yaoling Yang}
\email{yaolingyang@cuhk.edu.hk}
\affiliation{Department of Mechanical and Automation Engineering, 
The Chinese University of Hong Kong, Shatin, Hong Kong SAR, China}
\author{Paolo Bordone}
\email{paolo.bordone@unimore.it}
\affiliation{Dipartimento di Scienze Fisiche, Informatiche e Matematiche, Universit\`{a} di Modena e Reggio Emilia, I-41125 Modena, Italy}
\affiliation{Centro S3, CNR-Istituto di Nanoscienze, I-41125 Modena, Italy}
\author{Matteo G. A. Paris}
\email{matteo.paris@fisica.unimi.it}
\affiliation{Dipartimento di Fisica, Universit\`{a} di Milano, I-20133 Milano, Italy}
\author{Abolfazl Bayat}
\email{abolfazl.bayat@uestc.edu.cn}
\affiliation{Institute of Fundamental and Frontier Sciences, University of Electronic Sciences and Technology of China, Chengdu 611731, China}
\affiliation{Key Laboratory of Quantum Physics and Photonic Quantum Information, Ministry of Education, University of Electronic Science and Technology of China, Chengdu 611731, China}
\affiliation{Shimmer Center, Tianfu Jiangxi Laboratory, Chengdu 641419, China}


\date{\today}
\begin{abstract}
Metrology of defects (or impurities) is a major challenge in quantum sensing, with implications for material science, nano-scale metrology, and the development of high-precision quantum devices. In this work, we employ single and multi excitation sensors for the metrology of defects, with the system Hamiltonian encoding the structure of the sensor and the defect treated as a perturbation potential. We evaluate the quantum Fisher information of the ground state of the system to quantify the metrological performance of the sensors in locating a defect and in estimating its coupling strength. In doing so, we identify two distinct sensing regimes, determined by the ratio between the defect screening parameter and the inter-sensor site spacing, and establish the scaling behavior for the resulting estimation precision. Furthermore, we analyze the joint estimation of defect position and strength as well as the performance of two-excitation sensors, demonstrating how the number of excitations and their interactions influence sensitivity. Our results provide tools for optimizing defect detection with quantum sensors, with potential applications in solid-state quantum technologies.
\end{abstract}
\maketitle

\section{Introduction}
\label{sec:Introduction}

Defects in a quantum system are any imperfection that disrupts the regular order of the system. They typically arise from displaced or missing atoms or from the introduction of impurities into an otherwise pure lattice. From one side, defects may be an obstacle to building functional quantum devices, e.g. through inducing localization which prevents quantum transport~\cite{hundertmark2008short,segev2013anderson,abanin2019colloquium,nandkishore2015many} or imposing decoherence into neighboring qubits~\cite{maze2008electron,doherty2013nitrogen}. On the other hand, defects are a powerful resource for enabling quantum technologies, e.g., dopant-based qubits for computation~\cite{maurand2016cmos,Zwanenburg2013silicon,pla2012single,veldhorst2017silicon,he2019two} or nitrogen vacancies in diamonds for quantum sensing~\cite{aslam2023quantum,rembold2020introduction,balasubramanian2008nanoscale,dolde2011electric}. In either way, identifying the location of the defect and characterizing its impact on the neighboring environment is crucial~\cite{vaidya2023quantum,gottscholl2021spin,PhysRevResearch.5.043159,rizzato2023extending,casares2026quantum,jnane2026electron}.  
In a broad perspective, such characterization can be classified within the paradigm of Hamiltonian learning which has been pursued in two main directions~\cite{montenegro2025review}: (i) the static approach in which the probe is prepared in an equilibrium state, e.g. the ground state~\cite{zanardi2008quantum,raghunandan2018high,invernizzi2008optimal,mukhopadhyay2024modular,sarkar2022free,he2023stark,yousefjani2026exponentially,sarkar2025exponentially} or thermal state~\cite{rubio2021global,correa2015individual,mok2021optimal,mehboudi2019thermometry,campbell2018precision,jorgensen2022bayesian}; and (ii) the non-equilibrium approach in which the information is imprinted in the system through quantum dynamics induced by quantum quenches~\cite{chu2021dynamic,Boixo2007GeneralizedLimits,cavazzoni2025frequency,manshouri2025quantum,li2026nonequilibrium,mcdonald2020exponentially}, quantum walks~\cite{candeloro2020continuous, zatelli2020scattering,gianani2023multiparameter}, Floquet driving~\cite{mishra2021driving,mishra2022integrable,yang2026floquet,iemini2023floquet,Yousefjani2025PRApplied,fibger2026sensing}, quantum control~\cite{lang2015dynamical,pang2017optimal,schuff2020improving,xiao2022parameter,sekatski2017quantum}, particle shuttling~\cite{jnane2026electron} and sequential measurements~\cite{burgarth2015quantum, montenegro2022sequential,yang2023extractable,didi2022measurement}. 
The two directions are complementary in their resources and requirements. Static probing harnesses the probe size as a resource to enhance its precision while relying on (near-)equilibrium state preparation, without requiring time-resolved measurements. Dynamical probing, however, exploits time and probe size as resources to enhance its precision and avoids ground-state preparation at the cost of maintaining coherent evolution and time-resolved readout. In either case, the ultimate precision attainable with any given probe can be quantified by the quantum Fisher information (QFI) and its multiparameter generalization, the QFI matrix (QFIM)~\cite{paris2009quantum,liu2020quantum}. Here, we pursue the static strategy, in which the continuously parameterized position and coupling strength of the defect are encoded in the ground state of the sensor.

In this work, we employ a static approach, assuming a low-temperature, adiabatic scenario in which the probe, with single and two excitations, is prepared in its ground state, to identify the location and strength of a defect. We consider the probe to have a fixed spatial extension and control the number of sites among which the excitations can hop. A defect perturbs the local energy landscape, inducing a site-dependent energy shift that decays with distance and scales proportionally to the defect's strength. We systematically analyze the precision limits for estimating the defect's position as a single-parameter problem, alongside the joint estimation of its position and strength, which we formulate as a two-parameter problem. Based on the spatial extent of a defect's influence relative to the lattice spacing, we distinguish two distinct regimes: delocalized and localized defects. Delocalized defects can be detected with reasonable precision regardless of their position, with optimal sensitivity achieved when the defect is located at the center of the probe. In the localized regime, detection is possible only when the defect lies near one of the probe sites, and the maximum precision shifts to two symmetric points around the center of the probe. For a two-excitation probe, the behavior differs between attractive and repulsive interactions: when the defect is near the center of the probe, attractive interactions provide better sensitivity, whereas repulsive interactions are more useful for defects near the edges. Our analysis further reveals that, at fixed $\xi/a$, the number of sites in the probe is a key resource; the maximum of the QFI scales quadratically with $N$ in both regimes. This scaling stems from the increasing spatial resolution of a denser probe of fixed length, and it offers a significant advantage for the identification and characterization of defects~\cite{qp1,qp2,qp3,PhysRevA.94.053634,nokkala2016complex}.

The paper is structured as follows. In Sec.~\ref{sec:QS} we first present a brief review of quantum estimation theory, introducing the quantum Fisher information (QFI) and its relation to estimation precision, together with the corresponding tools used in the multiparameter case. In Sec.~\ref{sec:TM}, we then introduce the theoretical model adopted to describe both the defects and the sensors for the two cases of sensors containing either a single excitation or multiple excitations. Then, in Section~\ref{sec:S-ES}, we start focusing on the estimation of the position of the defect, and illustrate the sensing performance in two working regimes, defined by the ratio between the screening parameter of the defect potential and the distance between the sensor sites. We also study the finite size scaling of the QFI in the two regimes and the performance of the sensors in the joint estimation of position and coupling. In Section~\ref{sec:DPD_MPS} we perform a similar analysis for sensors involving two excitations. Section~\ref{sec:C} closes the paper with some concluding remarks. 

\section{Quantum sensing}
\label{sec:QS}
To study the metrological properties of a sensor, we assume operation at low temperature, and
in the adiabatic scenario, i.e., the insertion of the impurity leaves the system in its ground 
state. The ground state itself belongs to a class of pure states $|\psi_\lambda\rangle$ 
where $\lambda$
denotes one or more defect parameters.  Repeated measurements in the system provide information 
about the impurity, and the precision of any unbiased estimator of the defect parameters 
is bounded by the quantum Cram\'er-Rao theorem~\cite{paris2009quantum}. 
For strategies aimed at estimating a single parameter, the variance of any unbiased estimator is bounded by the single parameter QCR bound 
\begin{equation}
    \label{eq:qvar}
	\hbox{Var}\, x  \geq \frac{1}{M\,H(x)},
\end{equation}
where $M$ is the number of repeated measurements (the statistical scaling) and $H(x)$ is the 
quantum Fisher information (QFI) of the model. For a family of pure states, the QFI 
is given by 
\begin{equation}
    \label{eq:quantum_fisher_information_pure_states}
    H(x) = 4 \left( \langle \partial_x \psi_x \vert \partial_x \psi_x \rangle - \left| \langle \partial_x \psi_x\vert \psi_x \rangle\right|^2 \right)\,.
\end{equation}
The bound in Eq.~\eqref{eq:qvar}  is achievable, meaning that there exists an observable whose Fisher information is equal to the QFI.

If we are interested in more than one parameter, say the position $x$ and the strength of the impurity $h$, the sum of variances is bounded by 
\begin{align}
\hbox{Var}\, x + \hbox{Var}\, h \geq \frac{C_H}{M}
\end{align}
where $C_H$ is referred to as the {\em Holevo bound}, obtained by minimizing the sum of variances over all possible measurements, including collective measurements in multiple preparations of the sensor. \textcolor{black}{The Holevo bound may not be evaluated analytically, but it may be well approximated by exploiting the following chain of inequalities}
\begin{align}
\label{eq:bounds}
C_1 \leq C_H \leq C_1 + 2 \frac{s}{\sqrt{c}} \equiv C_2 \le 2 C_1\,,
\end{align}
where   
\begin{align}
C_1   = \hbox{Tr}[{\boldsymbol H}^{-1}(\lambda)] \quad
s   = 1/\hbox{det}[{\boldsymbol H}(\lambda)] \quad
c   = 1/\hbox{det}[{\boldsymbol U}(\lambda)]\,.
\end{align}
For the specific case of 2 parameters, say $\lambda=\{x,h\}$, the above definition leads to $C_1 = ( H_{11} +H_{22}) s$. The two matrices ${\boldsymbol H}(\lambda)$ and ${\boldsymbol U}(\lambda)$ are usually referred to as the quantum Fisher information matrix (QFIM) and the Uhlmann matrix, respectively. For pure statistical models, they may be evaluated as 
\begin{align}
    \label{eq:H U pure}
    H_{nm} &= 4 \hbox{Re}\left[ \langle \partial_n \psi_\lambda \vert \partial_m \psi_\lambda \rangle - \langle \partial_n \psi_\lambda\vert \psi_\lambda \rangle \langle \psi_\lambda\vert \partial_m \psi_\lambda \rangle  \right] \\
    U_{nm} &= 4 \hbox{Im}\left[ \langle \partial_n \psi_\lambda \vert \partial_m \psi_\lambda \rangle - \langle \partial_n \psi_\lambda\vert \psi_\lambda \rangle \langle \psi_\lambda\vert \partial_m \psi_\lambda \rangle  \right]\notag \,,
\end{align}
with $n,m=x,h$. 
The quantity $s$ is called the {\em sloppiness} of the statistical model. Large values of $s$ indicate poor parameter encoding, which means that the state of the sensor depends primarily on a single effective parameter, a combination of $x$ and $h$, rather than on each parameter individually~\cite{yang2025overcoming,mihailescu2025metrological,bizzarri2025controlling,sharma2025mitigating}. The quantity $c$, on the other hand, represents the {\em compatibility} of the model and measures the inverse of the average value of the commutator between the optimal observables for each parameter (i.e., those that maximize the QFI for a single parameter). Small values of $c$ signal the presence of quantum noise arising from the joint measurement of incompatible observables. Directly from Eq.\eqref{eq:H U pure} it is straightforward to demonstrate that a vanishing Uhlmann matrix $U=0$, e.g., in the case of a real wave function $\ket{\psi_\lambda}$,  leads to $\det(U)=0$ which directly implies $C_{1} = C_{H}$, leading to a compatible multiparameter quantum metrology. \textcolor{black}{The vanishing of the Uhlmann matrix ensures that the multiparameter Quantum Cramér-Rao bound is saturable, allowing all parameters to be simultaneously estimated with maximum theoretical precision using a single optimal measurement scheme.}

\section{Structure of the probe}
\label{sec:TM}

\begin{figure}[ht!]
\centering
\includegraphics[width=0.8\columnwidth]{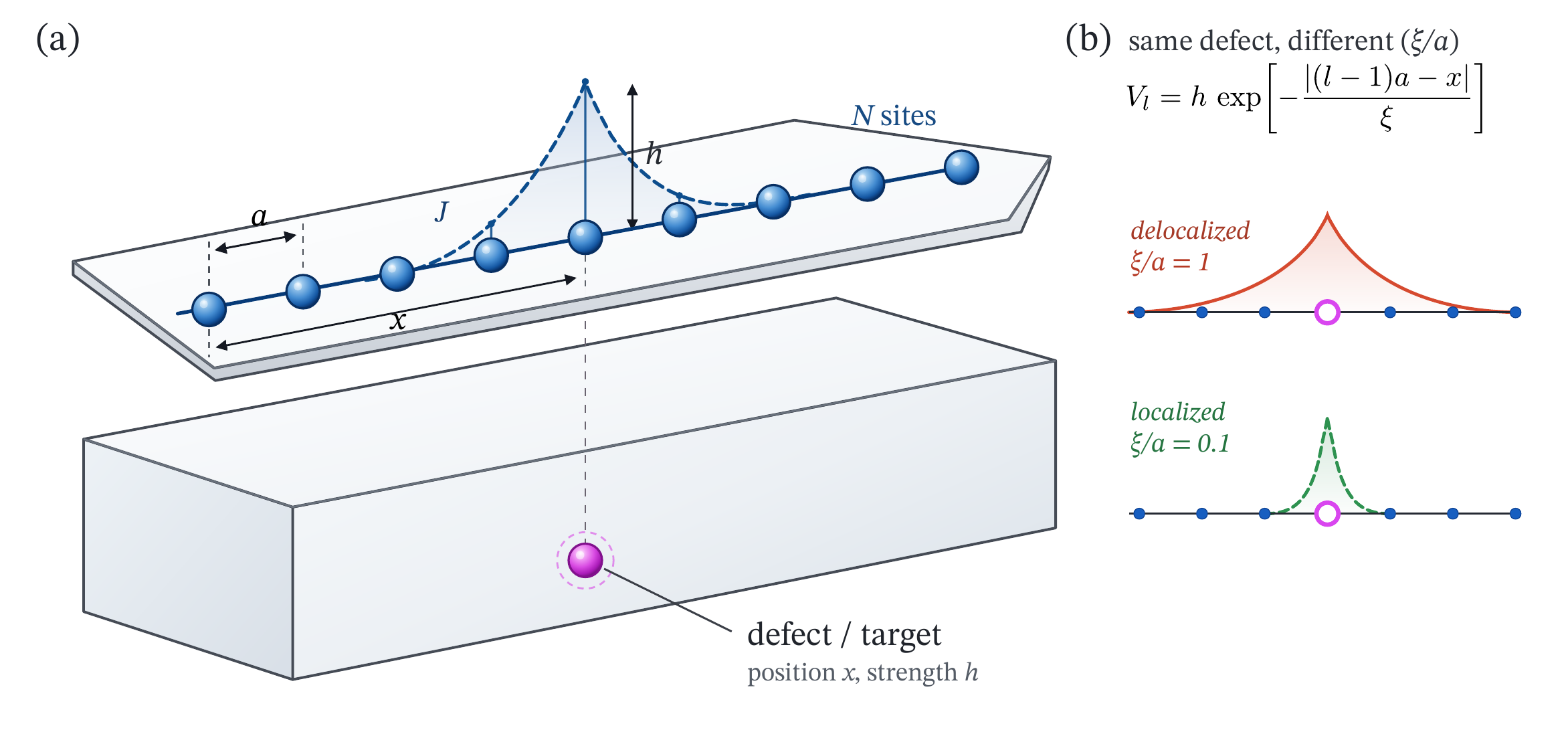}
\caption{\justifying \textbf{(a)} Schematic of the defect-sensing model. A one-dimensional
probe consists of $N$ sites (blue spheres), separated by a distance $a$ and coupled by
nearest-neighbor tunneling $J$. A single defect (magenta sphere) at position $x$, with
strength $h$, induces the on-site potential
$V_l=h\exp[-\lvert(l-1)a-x\rvert/\xi]$. \textbf{(b)} Profiles of the same defect for $\xi/a=1$ (delocalized regime, red solid curve) and $\xi/a=0.1$ (localized regime, green dashed curve).}
\label{fig:model_sensor}
\end{figure}

\textcolor{black}{As a first model of the quantum probe, we consider a single-excitation system. In this configuration, the probe is modeled as a one-dimensional spin chain where excitations tunnel between nearest-neighbor sites separated by a spatial distance $a$. In the single-excitation subspace, the Hamiltonian defining the probe takes the form}
\begin{align}
    \label{eq:sensor_Hamiltonian}
    \hat{\mathcal{H}}_{0,1e} = -J \sum_{l=1}^{N-1} \big( \vert l \rangle \langle l+1 \vert + h.c. \big),
\end{align}
in which $J$ represents the tunneling strength between adjacent sites and $N$ is the total number of sites.  The presence of the defect leads to an additional perturbative on-site potential, $\hat{\mathcal{V}}$, such that the overall Hamiltonian reads as follows
\begin{align}
    \label{eq:total_Hamiltonian}
    \hat{\mathcal{H}} = \hat{\mathcal{H}}_{0,1e} + \hat{\mathcal{V}}(h,x,a,\xi),
\end{align}
where we assume that the potential $\hat{\mathcal{V}}(h,x,a,\xi)$ is a screened potential of the form
\begin{align}
    \label{eq:defect_potential}
     \hat{\mathcal{V}}(h,x,a,\xi)  = h\, \sum_{l=1}^N  g_l (x,a,\xi) \, \ket{l}\bra{l},\, \quad \mathrm{where } \quad
      g_l (x,a,\xi)  = e^{-\frac{\abs{(l-1)a-x}}{\xi}}. 
\end{align}
The potential, which is diagonal in the position basis, is defined by three different parameters not related to the sensor: the defect position $x$, the coupling strength $h$, and the screening parameter $\xi$, which characterizes the spatial range of the defect-sensor interaction. Motivated by characteristic atomic electron density distributions~\cite{hu2007fitting,santoro2007complex}, this profile is adopted to ensure a realistic metrological characterization of defects in solid state systems. Moreover, the exponentially decaying defect–probe interaction potential may also serve as a natural benchmark potential, as a sensor resolving an exponential profile in the form of Eq.\eqref{eq:defect_potential}  is expected to also detect other decaying profiles, such as power-law potentials, establishing a fundamental benchmark for localized defect sensing. The potential has an additional dependence based on the structure of the probe itself: the separation of its sites $a$. To characterize the effect of this parameter in the metrological investigation of external defects, we assume that our probe has a fixed length (unit) and therefore the distance between the sites can be controlled by the number of sites as $a=1/(N-1)$. Although the defect parameter $\xi$ is fixed for a given defect, the spatial parameter $a=1/(N-1)$ depends on the number of sites and thus can be engineered at will. Consequently, varying $a$ enables full exploration of different potential configurations, whose profile in Eq.\eqref{eq:defect_potential} depends directly on the ratio $\xi/a$. In the following, we will analyze different sensor working regimes, characterized by the ratio between the screening parameter $\xi$ and the sensor spacing $a$, and perform a scaling analysis as a function of $N$. {Particularly, we will firstly focus on the estimation of the sole position of the defect $x$ and then on the joint estimation of the position $x$ and the strength of the defect $h$ to characterize both the single and multi parameter estimation performances of the sensor.} The structure of the sensor, together with the different potential profiles defined by the ratio $\xi/a$, is illustrated in Fig.~\ref{fig:model_sensor}. 
We also analyze the quantum sensor in the two-excitation subspace. Using the ordered position basis $\{\ket{j,k}:1\leq j<k\leq N\}$, where $\ket{j,k}$ denotes the state with one excitation at site $j$ and the other at site $k$, the Hamiltonian can be written as
\begin{equation}
\hat{\mathcal{H}} = \hat{\mathcal{H}}_{0,2e} + \hat{\mathcal{V}}(h,x,a,\xi),
\label{eq:two_excitation_hamiltonian}
\end{equation}
where
\begin{equation}
\hat{\mathcal{H}}_{0,2e} = -J \Bigg[ \sum_{j=1}^{N-2} \sum_{k=j+2}^{N} \ket{j+1,k}\bra{j,k} + \sum_{j=1}^{N-2} \sum_{k=j+1}^{N-1} \ket{j,k+1}\bra{j,k} + \mathrm{H.c.} \Bigg] + 2\Delta \sum_{j=1}^{N-1} \ket{j,j+1}\bra{j,j+1},
\label{eq:two_excitation_free_hamiltonian}
\end{equation}
and the defect potential is
\begin{equation}
\hat{\mathcal{V}}(h,x,a,\xi) = h \sum_{j=1}^{N-1} \sum_{k=j+1}^{N} \left[ g_j(x,a,\xi) + g_k(x,a,\xi) \right] \ket{j,k}\bra{j,k}.
\label{eq:two_excitation_defect_potential}
\end{equation}
Here, the two sums in the hopping term describe the allowed motion of the two excitations, while the last term in $\hat{\mathcal{H}}_{0,2e}$ gives the interaction energy when the excitations occupy neighboring sites. The defect potential is diagonal in the position basis and contains the local potential contributions experienced by both excitations.
After defining the Hamiltonian, a clarification must be made regarding its physical interpretation. In both single- and two-excitation regimes, the sensor Hamiltonian can describe either hopping particles on a lattice or excitations through a spin chain. For concreteness, we hereafter refer to these cases as single- and two-excitation sensors.

\section{Single-Excitation Sensor}
\label{sec:S-ES}

\textcolor{black}{In this Section, we assume that we work at low temperature, such that the probe is in the ground state, and we prepare the probe with a single excitation. At first, we address the metrological performances of the sensor in identifying the position of the defect $x$, and then we proceed 
with the analysis of the joint estimation of both the position $x$ and the strength $h$ of the defect.}

\subsection{Estimation of the position of the defect}
\label{subsec:DPD}
\begin{figure}[h!]
    \centering
    \parbox[b]{0.33\columnwidth}{
        \centering
        \textbf{(a)}\\
        \includegraphics[width=0.33\columnwidth]{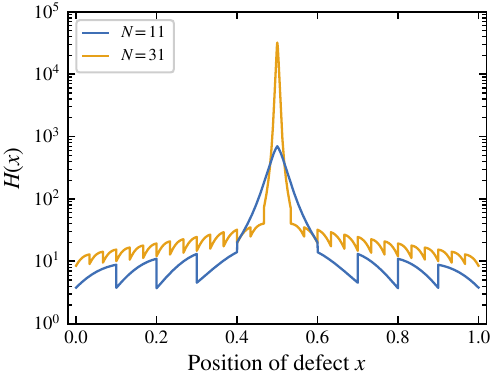}
    }
    \parbox[b]{0.33\columnwidth}{
        \centering
        \textbf{(b)}\\
        \includegraphics[width=0.33\columnwidth]{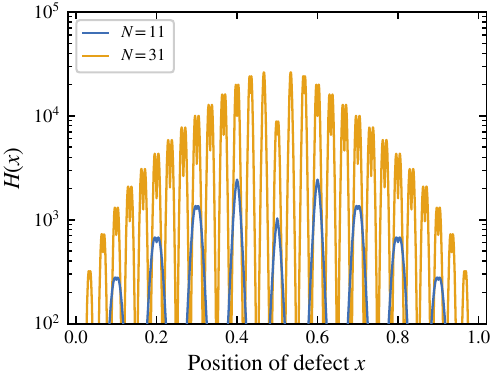}
    }
    \parbox[b]{0.33\columnwidth}{
        \centering
        \textbf{(c)}\\
        \includegraphics[width=0.33\columnwidth]{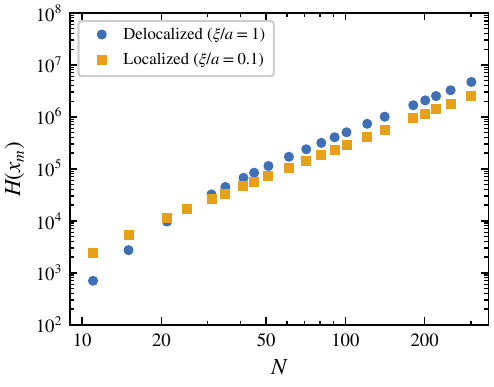}
    }
    \caption{\justifying 
    The quantum Fisher Information $H(x)$ (semi-log scale) governing the precision in the estimation of 
    the position of the defect and its scaling. Panel \textbf{(a)} refers to the delocalized regime ($\xi/a = 1$) and Panel \textbf{(b)} is for the localized regime ($\xi/a = 0.1$).
    The blue curves show the results for $N=11$ and the orange ones for $N=31$. 
    When the defect position coincides with the position of an element of the sensor the QFI typically shows a peak; in almost all cases, the closer the defect is to the center of the sensor, the higher the peak is. In all plots $h=J$. Panel \textbf{(c)}: Scaling of the maximum of the QFI for the two different regimes: $\xi/a=1$ (delocalized) and $\xi/a=0.1$ (localized) (different colors and shapes are associated to the different regimes).}
    \label{fig:xia}
\end{figure}

\textcolor{black}{As a first case study, we set $h=J$ and study the behavior of the QFI $H(x)$ for sensing the location $x$ of the defect. Depending on the ratio between the lattice distance $a$ and the screening parameter $\xi$, we may identify different sensing regimes, as the potential may involve fewer sensor sites (see Fig.~\ref{fig:model_sensor}). As it is apparent from Fig.~\ref{fig:xia}(a) and (b), two distinct regimes emerge: the delocalized regime $\xi/a \sim 1$ and the localized regime $\xi/a \simeq 0.1$.}
 We generalize this result in the multiparameter section when we obtain the QFI as a function of both the strength of the defect and its position.

In the {\em delocalized regime}, i.e., $\xi/a \sim 1$ (red curve in Fig.~\ref{fig:model_sensor}), the QFI $H(x)$ has the behavior shown in Fig.~\ref{fig:xia}(a). Several features can be observed: (i) QFI is always nonzero, showing that the position of the defect can always be identified; (ii) the QFI typically exhibits local maxima when the defect is positioned exactly at the sensor locations; (iii)  the global maximum of the QFI corresponds to the case when the defect lies exactly at the center of the sensor; and (iv) increasing the number of sensor sites $N$ not only results in a greater multiplicity of local maxima, but also leads to a substantial enhancement in the global maximum value of the QFI at the center of the chain. All these observations indicate that the probe is very effective in detecting the defect in the delocalized regime.

In the {\em localized regime}, i.e., $\xi/a \simeq 0.1$ (green curve in Fig.~\ref{fig:model_sensor}), new features emerge. In Fig.~\ref{fig:xia}(b) we plot the QFI as a function of $x$ when $\xi=0.1a$ for two different probes of length $N=11$ and $N=31$. As before, by increasing the number of sites in the sensor we see an increasing number of peaks in the QFI. However, unlike the delocalized case, in this regime the QFI becomes very small when the defect location is a bit away from the probe sites. This is due to the extreme localized impact of the defect which makes its impact observable only at the vicinity of the defect. In addition, the QFI exhibits two peaks located symmetrically around the central point, shifting the global maximum away from the midpoint.

The two regimes, discussed above, naturally suggest employing an adaptive estimation strategy. At first, one starts with a sensor with large $N$, namely small $a=1/(N-1)$, to guarantee that we operate in the delocalized regime. In this regime, the QFI is nonzero irrespective of the defect's location and thus some information can always be gained about the position of the defect. This information is then used to either suitably shift the position of the sensor itself to put the defect at the middle of the probe where the QFI is maximum or use a less dense sensor, with smaller $N$ and thus larger $a$, to operate in the localized regime such that the defect is located in the vicinity of one of the central sites in which the QFI is huge.

\begin{figure}[!ht]
    \centering
    \parbox[b]{0.33\columnwidth}{
        \centering
        \textbf{(a)}\\
        \includegraphics[width=\linewidth]{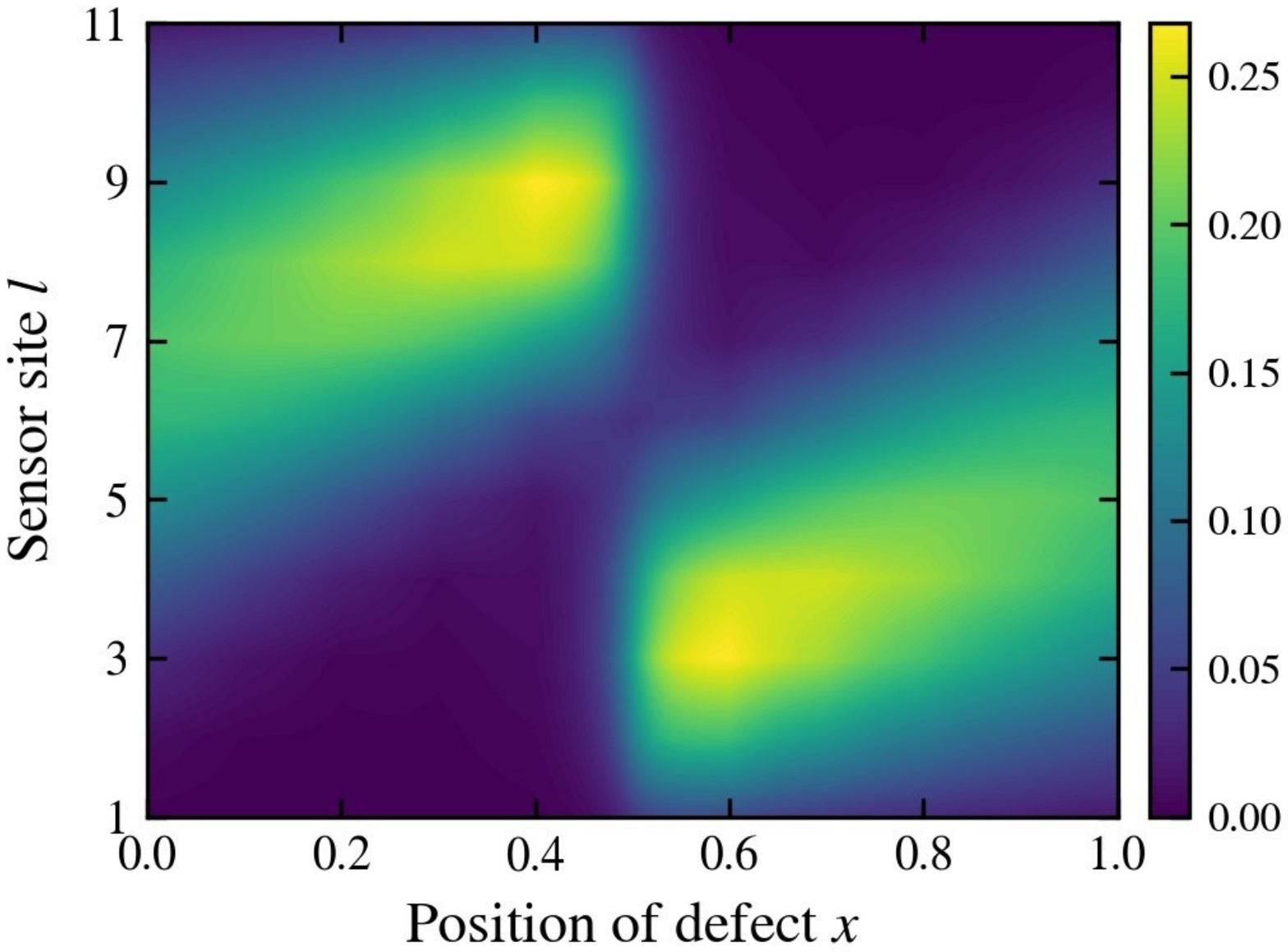}
    }
    \parbox[b]{0.33\columnwidth}{
        \centering
        \textbf{(b)}\\
        \includegraphics[width=\linewidth]{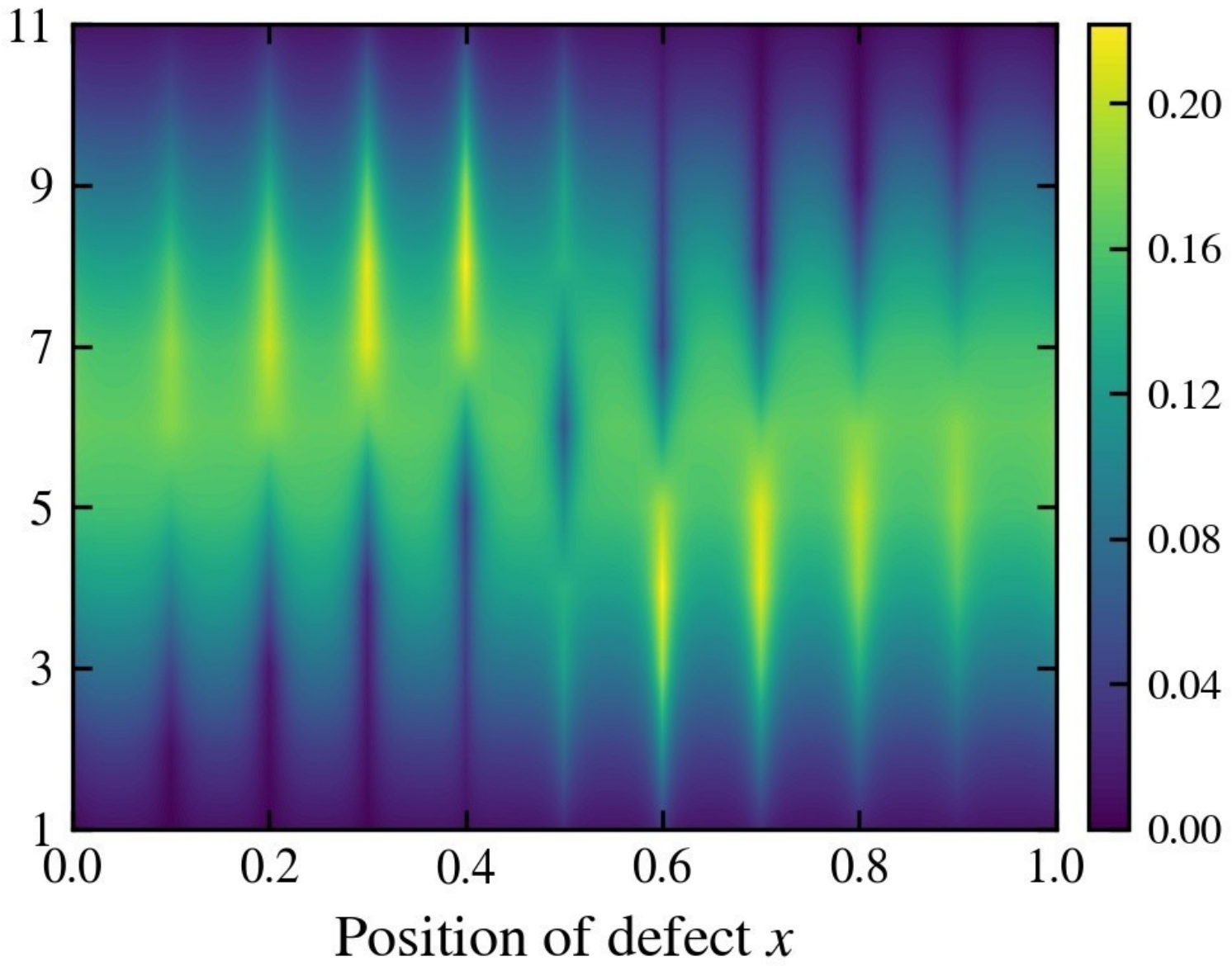}
    }
    \caption{\justifying 
    Heatmap of the probability density of the sensor $p_{l}(x)$ at low temperature and in the adiabatic scenario in the two different regimes: \textbf{(a)} delocalized regime ($\xi/a = 1$) and \textbf{(b)} localized regime ($\xi/a = 0.1$).
    In correspondence with an abrupt change of the density distribution in the sensor, the QFI shows a peak (see Fig.\ref{fig:xia}). All panels correspond to the case $N=11$ and $h=J$. 
    \label{fig:heatmap_1p}}
\end{figure}

Additionally, based on the above analysis, the number of sites $N$ emerges as a resource for estimating the location of the defect, as the maximum of the QFI $H(x)$ increases with $N$. In Fig.~\ref{fig:xia}(c), we plot the peak of the QFI that takes place at $x=x_m$ as a function of $N$ for two values of $\xi/a$ corresponding to the delocalized and localized regimes. As the figure shows, the QFI scales quadratically with $N$ as 
\begin{equation}
    H(x_m) \simeq b(\xi/a,x_m)\, N^2
\end{equation}
where the coefficient $b$ depends on the ratio $\xi/a$ and is different for different maxima $x_m$. The behavior of the QFI in all regimes can be understood by looking at the distribution of probability density 
\begin{equation}
    \label{eq:probability}
    p_l(x)=|\langle l|\psi_x\rangle|^2 \,,
\end{equation}
where $\ket{\psi_x}$ represents the wave function of the ground state when the defect is at the location $x$. In Figs.~\ref{fig:heatmap_1p}(a) and (b) we plot $p_l(x)$ as a function of defect location $x$ and probe's sites $l$ for delocalized and localized defects, respectively. The figure clearly shows that when the defect is located on one side of the probe, the excitation is localized on the other side. When the defect is located around the center of the sensor, the wave function is delocalized on both sides and shows the highest sensitivity to variation in $x$, which is captured through the peak of the QFI around the center of the probe. Interesting behavior can also be observed between the two different regimes. By decreasing the ratio $\xi/a$ from panel (a) to panel (b) in Fig.~\ref{fig:heatmap_1p}, we can see that the oscillations of the wave function become more pronounced. This is consistent with the oscillations that one can observe in the QFI shown in Figs.~\ref{fig:xia}(a) and (b).

\subsection{Multi-parameter estimation}
\label{subsec:Multiparameter-Estimation}
Let us now move to the joint estimation of the two parameters that define the defect-detection framework, namely the defect strength $h$ and its position $x$. The properties of the ground state of the Hamiltonian ensure that the Uhlmann matrix is null, i.e., $U=0$, which implies that no intrinsic quantum noise arises or equivalently, that $c$ in Eq.~\eqref{eq:bounds} diverges and $C_H = C_1$. In the following, we therefore analyze the behavior of the precision bound $C_H = C_1=\Tr[\left( H^{-1}(x,h) \right)]$ and of the sloppiness $s=1/\hbox{det}[H(x,h)]$ in the two metrological regimes identified for the estimation of the position of the defect alone.

\begin{figure}[!ht]
    \centering
    \parbox[b]{0.33\columnwidth}{
        \centering
        \textbf{(a)}\\
        \includegraphics[width=\linewidth]{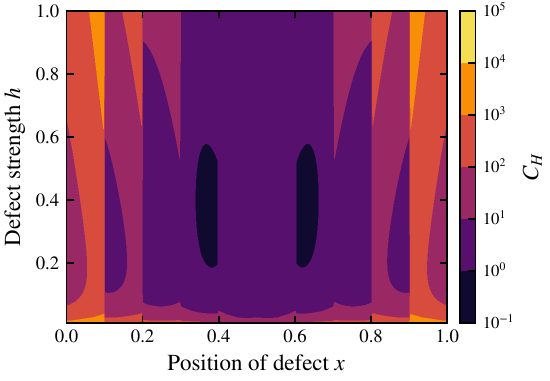}
    }
    \hspace{0.04\columnwidth}
    \parbox[b]{0.33\columnwidth}{
        \centering
        \textbf{(c)}\\
        \includegraphics[width=\linewidth]{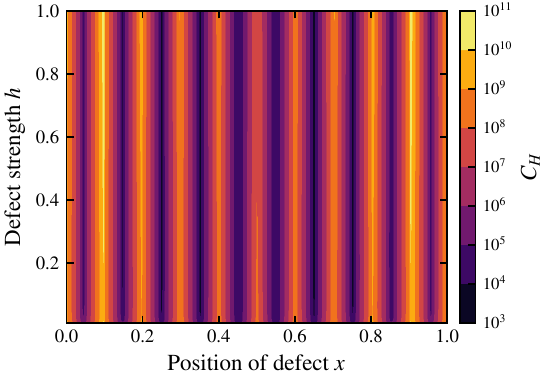}
    }

    \vspace{0.33cm} 

    \parbox[b]{0.33\columnwidth}{
        \centering
        \textbf{(b)}\\
        \includegraphics[width=\linewidth]{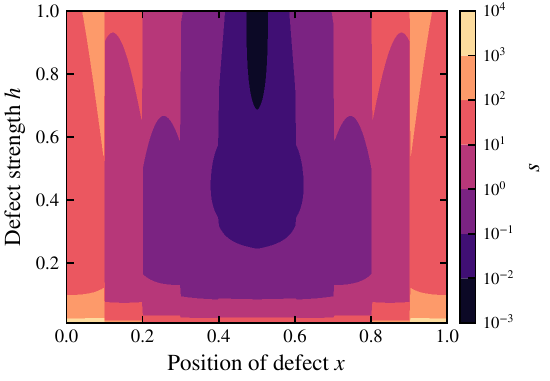}
    }
    \hspace{0.04\columnwidth}
    \parbox[b]{0.33\columnwidth}{
        \centering
        \textbf{(d)}\\
        \includegraphics[width=\linewidth]{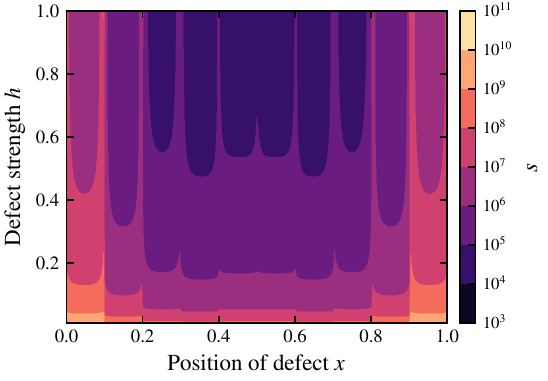}
    }

    \caption{\justifying 
    Heatmap of the precision bound $C_H = C_1 = \Tr[H^{-1}(x,h)]$ (top row) 
    and of the sloppiness $s = 1/\det(H(x,h))$ (bottom row) as a function of $x$ and $h$ for $N=11$. From left to right: delocalized regime ($\xi/a = 1$) [panels \textbf{(a)}, \textbf{(b)}] and localized regime ($\xi/a = 0.1$) [panels \textbf{(c)}, \textbf{(d)}].}
    \label{fig:TrDetxia}
\end{figure}

In the delocalized regime, $\xi/a \sim 1$, as is apparent in Fig.~\ref{fig:TrDetxia}(a) and (b), the values of $C_H$ and $s$ are closely related, which means that the bound on the sum of the variances of $h$ and $x$ is dominated by the determinant of the QFI matrix. For all values of $h$ the results follow those observed in Fig. \ref{fig:xia} (a) since the zone associated with the highest precision in the characterization of the defect (lowest value of $C_H$) is the one associated with the center of the sensor.
In the {\em localized regime}, $\xi/a \simeq 0.1$, the sensor is metrologically less efficient, as the precision bound is higher than in the delocalized regime. Indeed, looking at the scale of Fig.~\ref{fig:TrDetxia}(c) and (d), we can see that  $C_H$ and $s$ are higher than in the delocalized regime, and this affects the ultimate precision attainable in an estimation procedure. Nevertheless, in this regime, there are a higher number of metrologically interesting peaks of $C_H$, meaning that the estimation procedure may result in more robust results (i.e. the QFI is high in a broader zone). In both regimes, the determinant of the QFI matrix is therefore the main source of imprecision in the characterization of the defect, which is observed as a general property of the model in the single-excitation description.

\section{Two-Excitation Sensor}
\label{sec:DPD_MPS}

To generalize the single- and multi-parameter results of Sec.~\ref{sec:S-ES}, we move to the characterization of the defect using a two-excitation description of the sensor. Similarly to the single-excitation probe, we consider the probe to be in its ground state with the difference that the Hamiltonian is now given by Eq.~(\ref{eq:two_excitation_hamiltonian}). We begin our analysis by studying the QFI for the estimation of the position of the defect alone.

\subsection{Estimation of the position of the defect}

The metrological regimes identified in the previous sections remain valid for the characterization of the two-excitation sensor as well, and the following results are again categorized by the ratio $\xi/a$. 
To investigate the sensitivity of the sensor for detecting the defect's position $x$, we fix  $h=J$ and compute the quantum Fisher information $H(x)$. The interaction between the two excitations is taken to be  $\Delta \in \{-J,0, J\}$, see Eq.~\eqref{eq:two_excitation_hamiltonian}, to represent attractive, non-interacting and repulsive regimes.

\begin{figure}[!ht]
    \centering
    \parbox[b]{0.33\columnwidth}{
        \centering
        \textbf{(a)}\\
        \includegraphics[width=0.33\columnwidth]{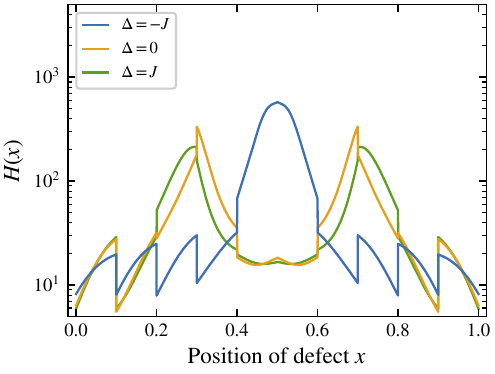}
    }
    \parbox[b]{0.33\columnwidth}{
        \centering
        \textbf{(b)}\\
        \includegraphics[width=0.33\columnwidth]{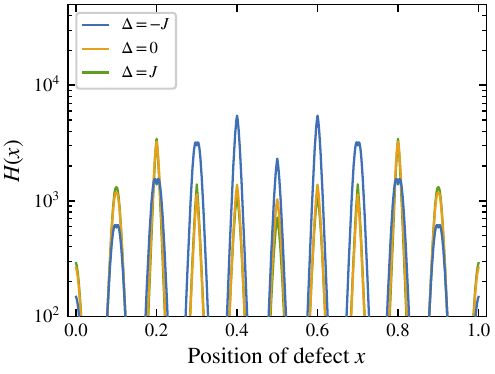}
    }
    \parbox[b]{0.33\columnwidth}{
        \centering
        \textbf{(c)}\\
        \includegraphics[width=0.33\columnwidth]{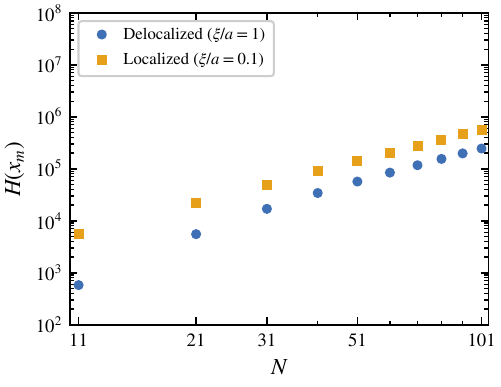}
    }
    
    \caption{\justifying 
    Semi-log plot of the quantum Fisher Information $H(x)$, governing the precision in the estimation of the position of the defect and its scaling. Panel \textbf{(a)} refers to the delocalized regime ($\xi/a = 1$) and Panel \textbf{(b)} is for the localized regime ($\xi/a = 0.1$).
    Different curve colors represent different values of the interaction parameter $\Delta$: 
    blue for $\Delta = -J$, orange for $\Delta = 0$, and green for $\Delta = J$. 
    Panels \textbf{(a)} and \textbf{(b)} correspond to $N = 11$ and $h = J$. Panel \textbf{(c)}: Scaling of the maximum QFI for the two regimes (distinct colors and shapes represent different regimes). All scaling data in \textbf{(c)} correspond to $\Delta = -J$, which yields the highest QFI maximum.}
    \label{fig:xia2p}
\end{figure}

In the {\em delocalized regime} ($\xi/a \sim 1$), we plot the QFI as a function of $x$ in Fig.~\ref{fig:xia2p}(a) for a probe of size $N=11$ and three different interaction strengths. Compared to the single excitation sensor, shown in Fig.~\ref{fig:xia}(a), the QFI of the two-excitation sensor shows some important differences depending on the value of $\Delta$. For $\Delta \geq 0$ (non-interacting or repulsive interaction), the QFI no longer reaches its maximum at the center of the sensor, but instead exhibits two symmetric peaks around the central point, while for $\Delta < 0$  (attractive interaction) the central peak is restored and the maximum of QFI is higher compared to the cases where $\Delta \geq 0$. The results clearly show that if the defect is located near the center of the probe, the attractive interaction ($\Delta<0$) shows better performance and if the defect is located towards the edges of the probe, then the repulsive interaction ($\Delta>0$) shows superior sensitivity.   This means that, in general, the interaction value $\Delta$ affects the performance of the probe, and its effect depends on the location of the defect, which is consistent with previous observations in Refs.~\cite{cavazzoni2024characterization,adani2024critical}. Another interesting feature is that the maximum QFI attainable by a two-excitation sensor can exceed that of a single-excitation sensor, indicating that the number of excitations is also a sensing resource. 

\begin{figure}[!ht]
    \centering
    \parbox[b]{0.33\columnwidth}{
        \centering
        \textbf{(a)}\\
        \includegraphics[width=\linewidth]{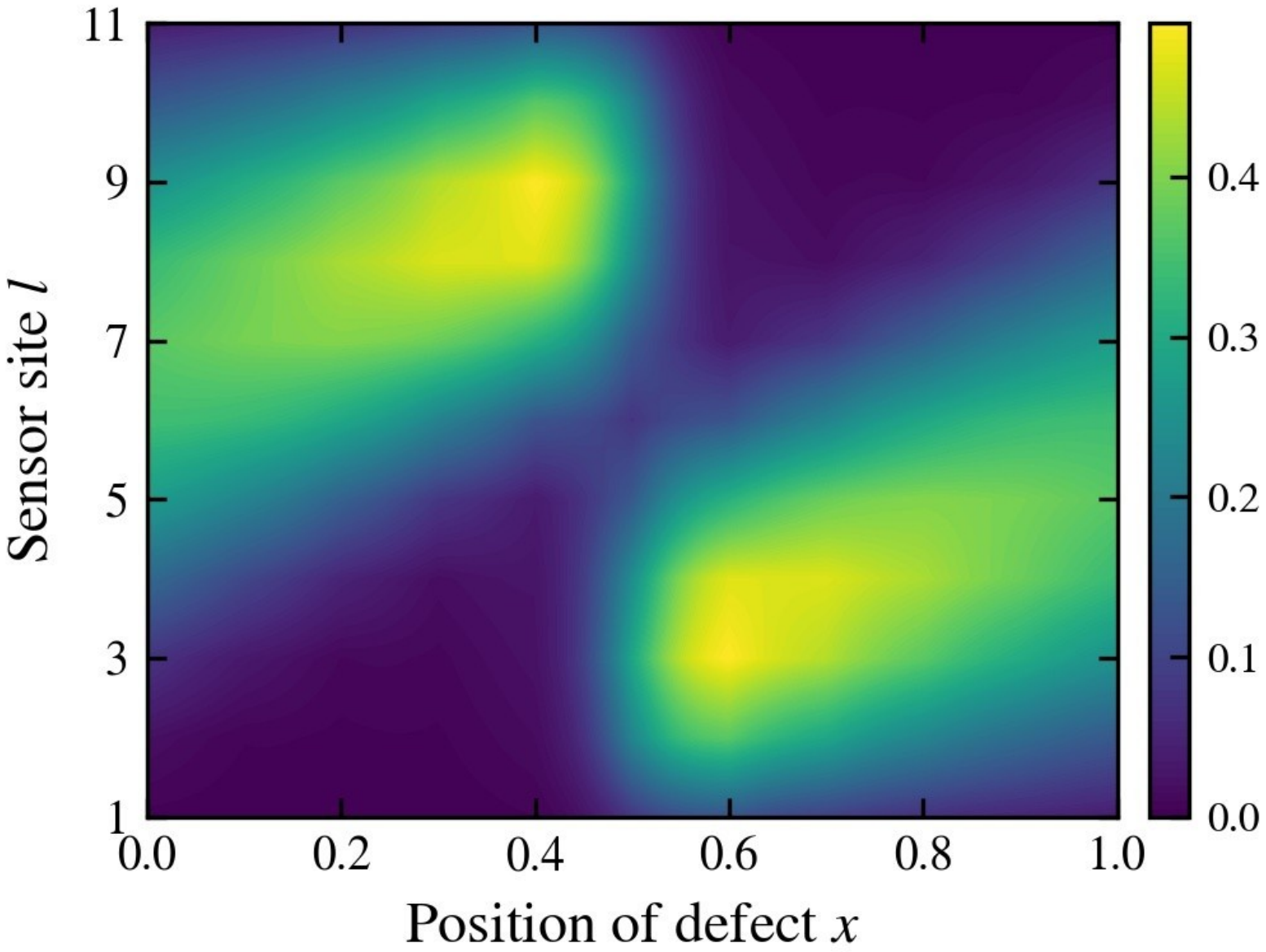}
    }
    \hspace{0.04\columnwidth}
    \parbox[b]{0.33\columnwidth}{
        \centering
        \textbf{(b)}\\
        \includegraphics[width=\linewidth]{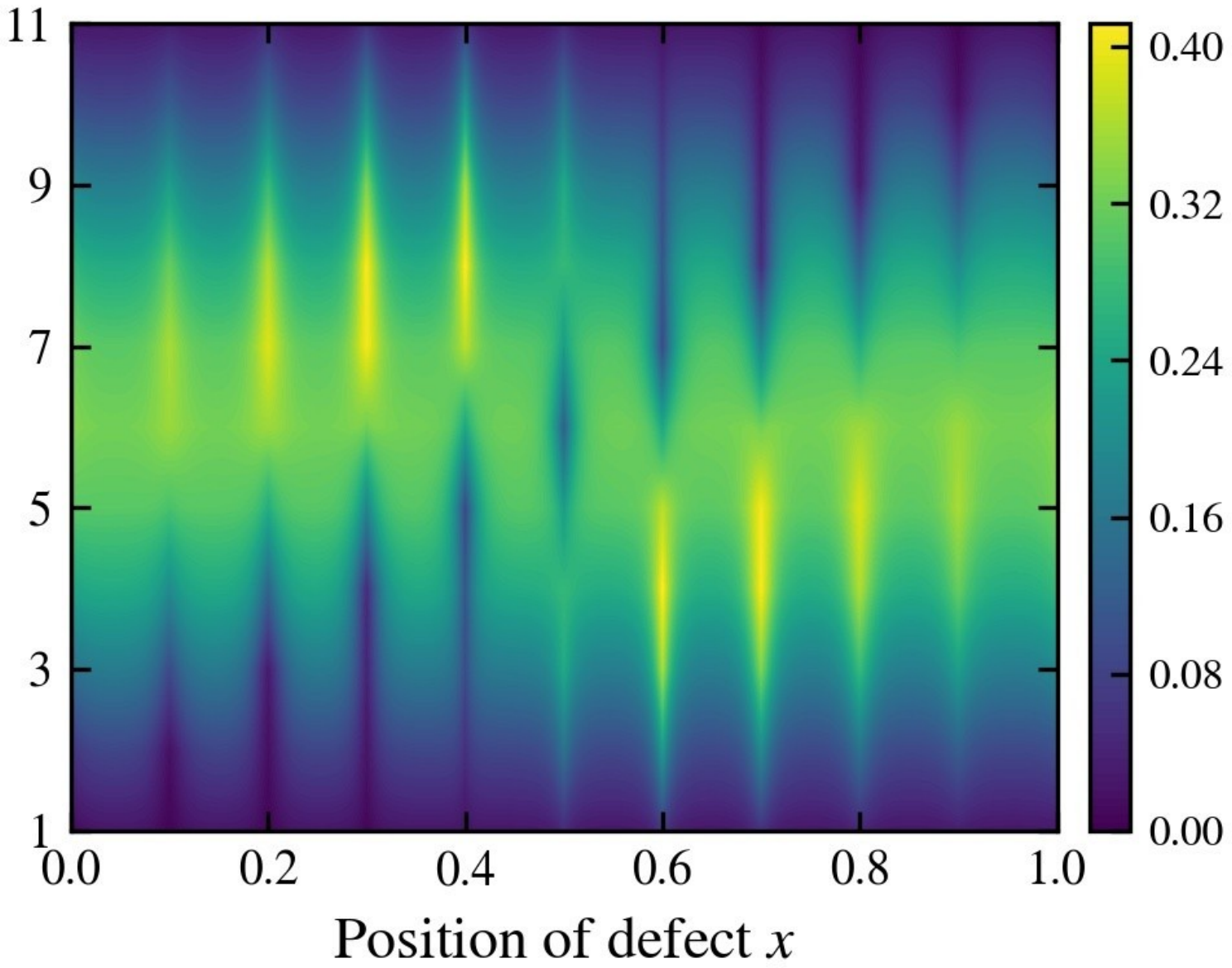}
    }
    \caption{\justifying 
    Heatmap of the probability density of the sensor $\rho_l(x)$ (see Eq.\eqref{eq:probability 2p}) in the ground state of the probe in two different regimes: \textbf{(a)} delocalized regime ($\xi/a = 1$) and \textbf{(b)} localized regime ($\xi/a = 0.1$).
    In correspondence with an abrupt change of the density distribution in the sensor, the QFI shows a peak (see Fig.\ref{fig:xia2p}). All panels correspond to the favorable interaction condition $\Delta = -J$ for the case $h = J$ and $N = 11$.} 
    \label{fig:heatmap_2p}
\end{figure}

In the {\em localized regime} ($\xi/a \simeq 0.1$), we plot the QFI as a function of $x$ in Fig.~\ref{fig:xia2p}(b) for a probe of size $N=11$ considering three different interaction values $\Delta$. As the figure shows, similar to the case of the single-excitation subspace, the QFI is nonzero only when the defect is located near the sites of the probe, and the maximum of the QFI is obtained when the defect is located at the sites $5$ and $7$, namely symmetrically around the central point.  Again, the effect of interaction depends on the location of the defect. If the defect is located at the center of the chain then attractive interaction ($\Delta<0$) shows better sensitivity, and if the defect is located towards the edges, repulsive interaction ($\Delta>0$) gives higher sensitivity. In addition, the values of the obtainable QFI are higher than in the case of the single-excitation sensor with a factor of almost $\sim 2$, suggesting that QFI increases linearly with the number of excitations, consistent with the results of Ref.~\cite{manshouri2025quantum}. To show the scaling of the QFI with respect to probe size, also for the two excitation sensors, in Fig.~\ref{fig:xia2p}(c), we plot the maximum QFI versus $N$, which again shows quadratic scaling as
\begin{equation}
    H(x_m) \simeq b_{2e}(\xi/a,x_m)\, N^2
\end{equation}
where the coefficient $b_{2e}$ increases when the ratio $\xi/a$ decreases and is different for different maxima $x_m$, see Fig.~\ref{fig:xia2p}(c). The behavior of the QFI again suggests that the same adaptive estimation strategy, described in Sec.\ref{sec:S-ES}, can be employed for estimating the position of the defect. Similar to the single-excitation scenario, it will be insightful to explore the wave function of the two-excitation sensor too. The joint probability density of finding the two excitations of the sensor at sites $l$ and $m$ (with $l<m$) is given by
\begin{equation}
    \label{eq: 2p density}
    \rho_{l,m}(x) = |\langle l,m|\psi_{x}\rangle|^2.
\end{equation}
There are two ways to explore the excitation density in the probe. In the first approach, one can explore the single-site occupation probability $\rho_l(x)$, which quantifies the probability of finding an excitation at site $l$, irrespective of the location of the other excitation. In the second approach, one can directly explore the joint probability density $\rho_{l,m}(x)$. Note that we use the notation $\rho_l(x)$ for the single-site occupation probability in the two-excitation probe to distinguish it from the probability density $p_l(x)$, see Eq.~(\ref{eq:probability}), in the single-excitation probe. Therefore, one can write
\begin{equation}
\label{eq:probability 2p}
\rho_l(x)= \sum_{w=l+1}^{N}\rho_{l,w}(x) + \sum_{r=1}^{l-1}\rho_{r,l}(x).
\end{equation}
In Figs.~\ref{fig:heatmap_2p}(a) and (b), we plot the probability distribution $\rho_l(x)$ as a function of probe site $l$ and defect position $x$. It is again evident that, for defect positions where the QFI peak is highest (according to Fig.~\ref{fig:xia2p}), the single-site occupation probability is delocalized around the center. Particularly in the delocalized regime, Fig.~\ref{fig:heatmap_2p}(a), this mainly occurs when the defect is at the center of the sensor, whereas in the localized regime it occurs when the impurity is near any sensor site; see Fig.~\ref{fig:heatmap_2p}(b). Compared to the single-excitation sensor, in the localized regime the maximum change in the probability density is larger, and consequently the maximum QFI is higher (see Fig.~\ref{fig:heatmap_1p} and Fig.~\ref{fig:heatmap_2p}).

\begin{figure}[!ht]
    \centering
    \parbox[b]{0.33\columnwidth}{
        \centering
        \textbf{(a)}\\
        \includegraphics[width=\linewidth]{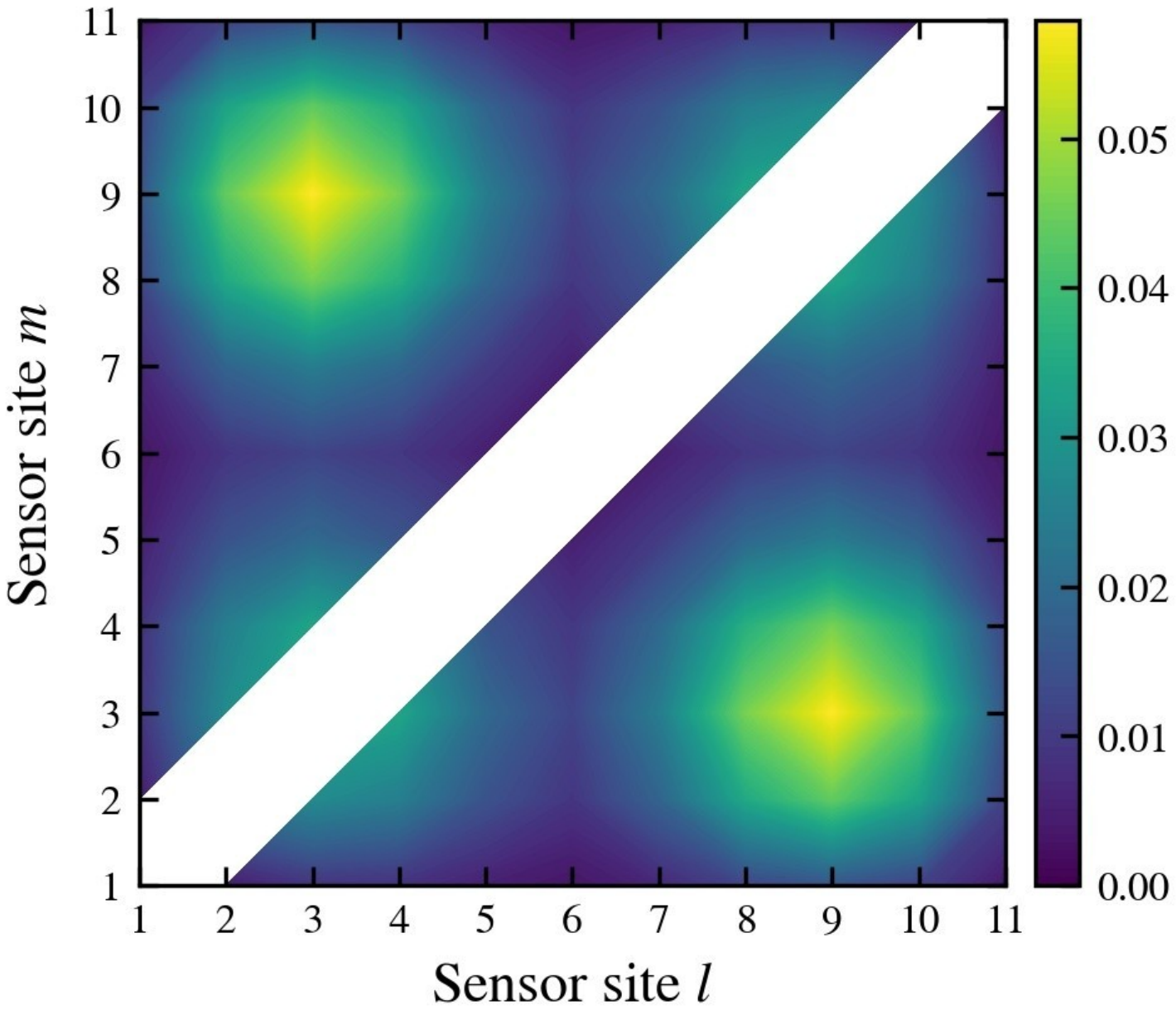}
    }
    \parbox[b]{0.33\columnwidth}{
        \centering
        \textbf{(b)}\\
        \includegraphics[width=\linewidth]{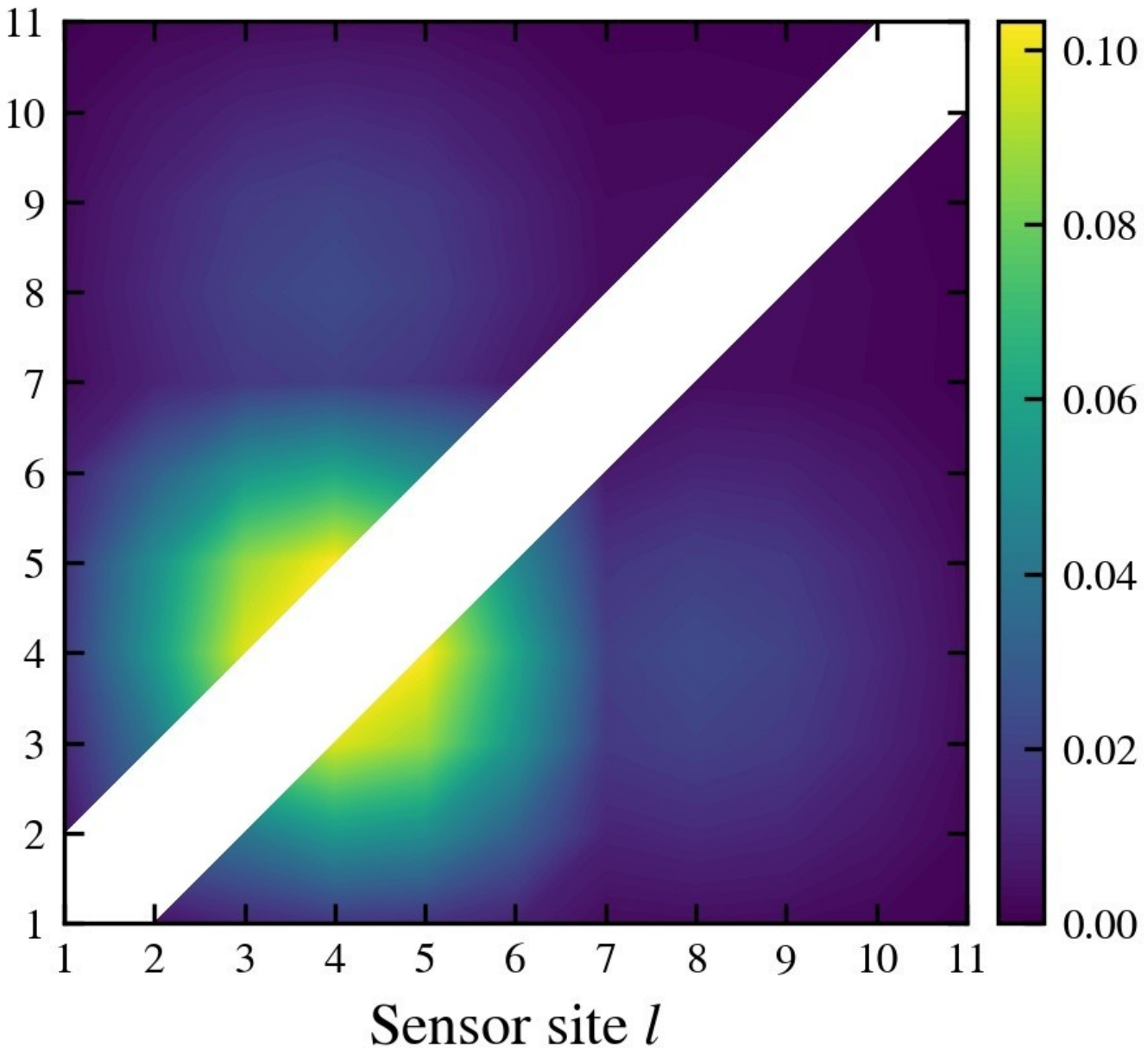}
    }
    \caption{\justifying 
    Heatmap of the joint probability density of the two excitations $\rho_{l,m}(x)$ in the ground state of the probe in two different regimes: \textbf{(a)} delocalized regime ($\xi/a = 1$) and \textbf{(b)} localized regime ($\xi/a = 0.1$).
    In each panel, $x$ is set to the value $x_{m}$ that maximizes the QFI for estimating the defect position, $H(x)$ (see Fig.\ref{fig:xia2p}). All panels correspond to the favorable interaction condition $\Delta = -J$ for the case $h = J$ and $N = 11$.}
    \label{fig:heatmap_2p_new}
\end{figure}

One can also directly look at the joint probability density $\rho_{l,m}(x)$. The results are shown in Figs.~\ref{fig:heatmap_2p_new}(a) and (b). The results support the results presented in Fig.~\ref{fig:heatmap_1p}. As it is possible to notice from the scale of each panel of Fig.~\ref{fig:heatmap_2p_new}, as $\xi/a$ decreases, the peak value of $\rho_{l,m}$ increases substantially from $\sim 0.058$ in the delocalized regime to $\simeq 0.10$ in the localized regime. This is consistent with the enhancement of $H(x_{m})$ in the two regimes identified, and directly relates the behavior of the QFI with the density of the system $\rho_{l,m}(x)$.

\subsection{Multi-parameter estimation}
\label{sec:MPE_MPS}

After analyzing the QFI associated with the impurity's position in the two-excitation formalism, we return to the joint estimation of the two quantities that fully define the defect-detection paradigm and characterize the metrological properties of the sensor and the impurity, namely $h$ and $x$. As in the single-excitation case, we also study the sensor properties in the equilibrium regime ($\ket{\psi_{\lambda}}=\ket{g.s.}$) within the two-excitation formalism. 

\begin{figure}[!ht]
    \centering
    \parbox[b]{0.33\columnwidth}{
        \centering
        \textbf{(a)}\\
        \includegraphics[width=\linewidth]{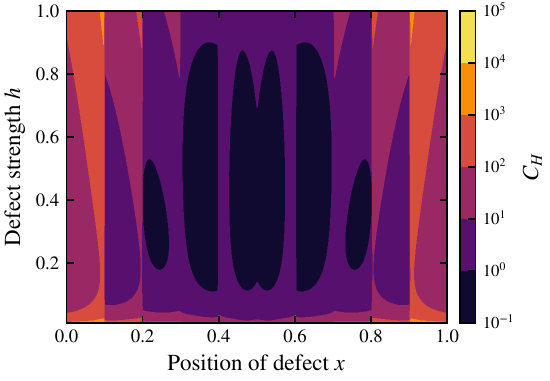}
    }
    \hspace{0.04\columnwidth}
    \parbox[b]{0.33\columnwidth}{
        \centering
        \textbf{(c)}\\
        \includegraphics[width=\linewidth]{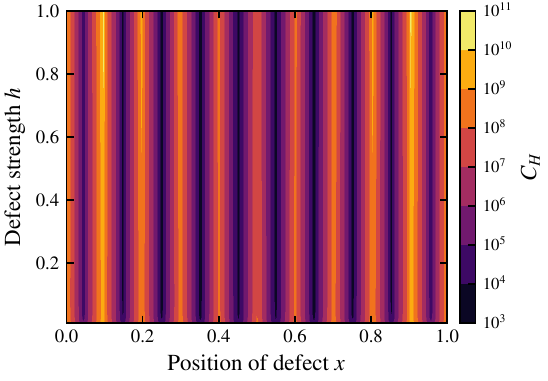}
    }

    \vspace{0.33cm} 

    \parbox[b]{0.33\columnwidth}{
        \centering
        \textbf{(b)}\\
        \includegraphics[width=\linewidth]{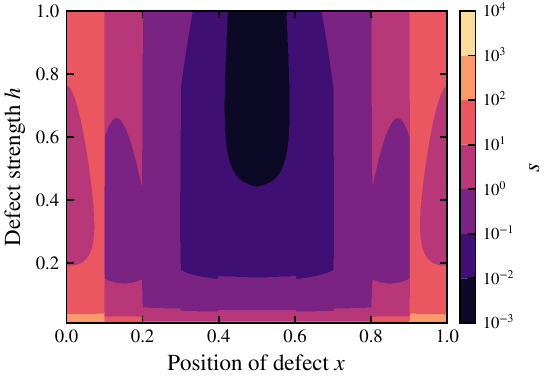}
    }
    \hspace{0.04\columnwidth}
    \parbox[b]{0.33\columnwidth}{
        \centering
        \textbf{(d)}\\
        \includegraphics[width=\linewidth]{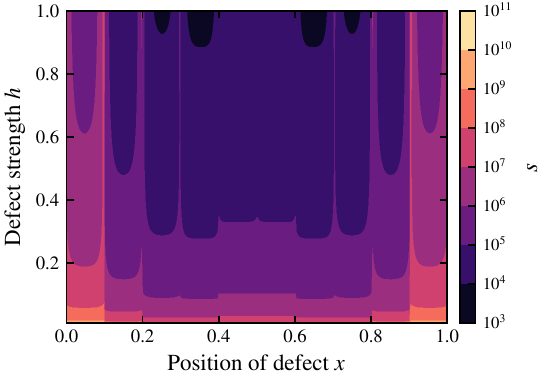}
    }
    \caption{\justifying 
    Heatmap of the trace of the inverse quantum Fisher information matrix (top row) 
    and of the inverse of the determinant (bottom row). 
    The results are related to the wavefunction $\ket{\psi_{x}}$ at low temperature and in the adiabatic scenario for $N = 11$. 
    From left to right: delocalized regime ($\xi/a = 1$) [panels \textbf{(a)}, \textbf{(b)}] and localized regime ($\xi/a = 0.1$) [panels \textbf{(c)}, \textbf{(d)}].}
    \label{fig:TrDetxia2p}
\end{figure}

The determinant of the QFI matrix is non-null, while the Uhlmann curvature matrix is again null, meaning that the simultaneous estimation of $h$ and $x$ is possible and that the two parameters are compatible, namely $c\rightarrow\infty$ in Eq.\eqref{eq:bounds}. As a result, the characterization of the sensor's metrological properties is given by the trace of the inverse of the QFI matrix, which equals the Holevo bound $C_H$, and by the sloppiness $s$. In the case of single-parameter estimation with a two-excitation probe, the attractive interaction (i.e. $\Delta<0$) results in a larger maximum QFI, see Fig.~\ref{fig:xia2p}. Therefore, without loss of generality, for the multi-parameter case, we only consider attractive interaction with $\Delta=-J$. 

In Figs.~\ref{fig:TrDetxia2p}(a)-(b), we plot $C_H$ and $s$ as a function of $h$ and $x$ for the delocalized regime with $\xi/a \sim 1$. As the figure shows, the two quantities show correlated behavior, which indicates that the sum of the variances of $x$ and $h$, described by $C_H$, is mainly determined by the determinant of the QFI matrix, described by the parameter $s$. In Figs.~\ref{fig:TrDetxia2p}(c)-(d), we plot $C_H$ and $s$ as a function of $h$ and $x$ for the localized regime with $\xi/a \simeq 0.1$. In this case, the correlation between $C_H$ and $s$ becomes very clear, showing that the enhancement of the estimation uncertainty of $x$ and $h$, quantified by $C_H$, is mainly due to the sloppiness parameter $s$. It is worth noting that in both regimes the sloppiness parameter $s$ is smaller in the two-excitation case than in the single-excitation probe. This implies that increasing the number of excitations is indeed a valuable resource for multi-parameter estimation, as multiple excitations can overcome or mitigate the probe's potential sloppiness.

\section{Conclusions}
\label{sec:C}

In this paper, we have investigated quantum sensing of defects by analyzing the ground state of a probe initialized in either the single- or the two-excitation subspace. Assuming a low-temperature scenario, we consider the probe to be in its ground state, whose properties are affected by the defect as a perturbation of the sensor Hamiltonian. Our study focuses on a single defect located near the probe, whose screened potential is characterized by its position, its strength, and its screening parameter, the latter defining the spatial range of the defect–sensor interaction. For defect detection, two distinct problems were considered. First,   single-parameter estimation for detecting the position of the defect. Second, two-parameter estimation for estimating both the position and strength of the defect. In the former, the figure of merit for quantifying the performance of the probe is the QFI while in the latter we consider  QFIM together with the associated sloppiness and compatibility parameters.  

For the single-excitation probe, our analysis reveals several key findings: (i) the probe exhibits two distinct operational regimes, termed delocalized and localized, determined by the ratio $\xi/a$ between the defect's screening parameter and the sensor spacing; (ii) delocalized defects are detectable regardless of their position, with optimal sensitivity achieved when the defect is centered within the probe; (iii) localized defects are only detectable when positioned near a probe site, with the maximum precision shifting to two symmetric points around the probe center; and (iv) at fixed $\xi/a$, the QFI at its maximum, and hence the ultimate estimation precision, scales quadratically with the number of probe sites, with a prefactor that depends on $\xi/a$.
Beyond the single-parameter problem, we have addressed the joint estimation of the defect position and its strength. The ground state of the Hamiltonian is real, so the Uhlmann matrix vanishes identically,  and the multiparameter quantum Cramér–Rao bound is therefore saturable: the two parameters are compatible, no intrinsic quantum noise arises, and the Holevo bound reduces to the trace of the inverse QFIM. In both regimes, we find that precision is limited by the sloppiness parameter of the model, rather than by the incompatibility of the optimal observables. While the delocalized regime offers the best joint precision, around the center of the probe, the localized regime exhibits a larger number of metrologically useful peaks, albeit at the cost of reduced precision. These results identify the joint estimation as a multiparameter problem in which the sloppiness of the model, rather than quantum incompatibility, sets the ultimate limit.

In the two-excitation subspace, in addition to the above findings, we have shown that the interaction between the two excitations plays an important role. While an attractive interaction ($\Delta < 0$) makes the probe more sensitive to defects in the central part of the probe, a repulsive interaction ($\Delta > 0$) is more suitable for detecting defects near the edges. The maximum QFI attainable by the two-excitation sensor can exceed that of the single-excitation sensor, and in the localized regime the enhancement is approximately a factor of two, consistent with a linear increase of the QFI with the number of excitations. The quadratic scaling is preserved, with a prefactor that increases as the ratio $\xi/a$ decreases. The multiparameter analysis for the two-excitation case reveals that the Uhlmann matrix remains null, so that the two parameters are jointly estimable, and the sloppiness is systematically smaller than in the single-excitation probe. This demonstrates that increasing the number of excitations is itself a valuable resource for multiparameter estimation, paving the way for the use of probes with a higher number of excitations in future sensor designs.

All these features, in both the single- and the two-excitation cases, are interpretable in terms of the spatial structure of the sensor's wavefunction, and they motivate an adaptive estimation strategy for defect sensing in which the lattice spacing $a$, and consequently the number of sites $N$, can be tuned to exploit different probe regimes. Initially, a densely packed sensor (small $a$, hence large $N$) can be employed to operate in the delocalized regime, enabling defect detection with reasonable precision irrespective of the defect's location. Once the defect's approximate position is known, a less dense probe operating in the localized regime can be used to achieve significantly enhanced measurement precision, provided the defect is placed in the vicinity of one of the central sites. This two-stage strategy balances a broad initial search with a subsequent high-precision localization, and it exploits the quadratic scaling of the QFI with $N$ as the primary resource.

These findings have direct relevance to the detection of charge impurities in quantum dot arrays, the characterization of magnetic defects in spin chains, and the assessment of disorder in materials used for quantum computing platforms. The generalization from single- to two-excitation sensors also illustrates how the excitation number can be used as a controlled resource for both single- and multiparameter estimation, suggesting that future sensor design could benefit from exploring probes with a higher number of excitations. Our study can be generalized further for detecting multiple impurities, which is left for a future project.

\begin{acknowledgments}
AB acknowledges support from the National Natural Science Foundation of China (grants No. W2541020, No. 12274059, No. 12574528 and No. 1251101297). This work has been done under the auspices of GNFM-INdAM and has been partially supported by MUR and EU through the projects PRIN22-2022T25TR3-RISQUE and PRIN22-PNRR-P202222WBL-QWEST. 
\end{acknowledgments}

\section*{Data Availability and Specifications}
No new data were generated or analyzed in this study. All results can be reproduced from the equations and methods presented in the manuscript. 

\bibliography{qwfqs.bib}

\end{document}